\documentstyle[graphicx,lscape,times]{mn}

\newif\ifAMStwofonts

\ifoldfss
  \ifCUPmtlplainloaded \else
    \NewTextAlphabet{textbfit} {cmbxti10} {}
    \NewTextAlphabet{textbfss} {cmssbx10} {}
    \NewMathAlphabet{mathbfit} {cmbxti10} {} 
    \NewMathAlphabet{mathbfss} {cmssbx10} {} 
  \fi
  \ifAMStwofonts
    \ifCUPmtlplainloaded \else
      \NewSymbolFont{upmath} {eurm10}
      \NewSymbolFont{AMSa} {msam10}
      \NewMathSymbol{\upi}     {0}{upmath}{19}
      \NewMathSymbol{\umu}     {0}{upmath}{16}
      \NewMathSymbol{\upartial}{0}{upmath}{40}
      \NewMathSymbol{\leqslant}{3}{AMSa}{36}
      \NewMathSymbol{\geqslant}{3}{AMSa}{3E}

       \let\le=\leqslant
       \let\ge=\geqslant
    \fi
  \fi
\fi 

\ifnfssone
  \newmathalphabet{\mathit}
  \addtoversion{normal}{\mathit}{cmr}{m}{it}
  \addtoversion{bold}{\mathit}{cmr}{bx}{it}
  \newmathalphabet{\mathbfit} 
  \addtoversion{normal}{\mathbfit}{cmr}{bx}{it}
  \addtoversion{bold}{\mathbfit}{cmr}{bx}{it}
  \newmathalphabet{\mathbfss} 
  \addtoversion{normal}{\mathbfss}{cmss}{bx}{n}
  \addtoversion{bold}{\mathbfss}{cmss}{bx}{n}
  \ifAMStwofonts
    \ifCUPmtlplainloaded \else
      \UseAMStwoboldmath
      \makeatletter
      \new@mathgroup\upmath@group
      \define@mathgroup\mv@normal\upmath@group{eur}{m}{n}
      \define@mathgroup\mv@bold\upmath@group{eur}{b}{n}
      \edef\UPM{\hexnumber\upmath@group}
      \new@mathgroup\amsa@group
      \define@mathgroup\mv@normal\amsa@group{msa}{m}{n}
      \define@mathgroup\mv@bold\amsa@group{msa}{m}{n}
      \edef\AMSa{\hexnumber\amsa@group}
      \makeatother
      \mathchardef\upi="0\UPM19
      \mathchardef\umu="0\UPM16
      \mathchardef\upartial="0\UPM40
      \mathchardef\leqslant="3\AMSa36
      \mathchardef\geqslant="3\AMSa3E

       \let\le=\leqslant
       \let\ge=\geqslant
    \fi
  \fi
\fi 

\ifnfsstwo
  \DeclareMathAlphabet{\mathbfit}{OT1}{cmr}{bx}{it}
  \SetMathAlphabet\mathbfit{bold}{OT1}{cmr}{bx}{it}
  \DeclareMathAlphabet{\mathbfss}{OT1}{cmss}{bx}{n}
  \SetMathAlphabet\mathbfss{bold}{OT1}{cmss}{bx}{n}
  \ifAMStwofonts
    \ifCUPmtlplainloaded \else
      \DeclareSymbolFont{UPM}{U}{eur}{m}{n}
      \SetSymbolFont{UPM}{bold}{U}{eur}{b}{n}
      \DeclareSymbolFont{AMSa}{U}{msa}{m}{n}
      \DeclareMathSymbol{\upi}{0}{UPM}{"19}
      \DeclareMathSymbol{\umu}{0}{UPM}{"16}
      \DeclareMathSymbol{\upartial}{0}{UPM}{"40}
      \DeclareMathSymbol{\leqslant}{3}{AMSa}{"36}
      \DeclareMathSymbol{\geqslant}{3}{AMSa}{"3E}

       \let\le=\leqslant
       \let\ge=\geqslant
    \fi
  \fi
\fi 

\ifCUPmtlplainloaded \else
  \ifAMStwofonts \else 
    \def\upi{\pi}
    \def\umu{\mu}
    \def\upartial{\partial}
  \fi
\fi

\title[Globular clusters in the outer halo]
  {Globular clusters and the formation of the outer Galactic halo}
\author[Sidney~van~den~Bergh \& A.~D.~Mackey]
  {Sidney~van~den~Bergh$^1$ and A.~D.~Mackey$^2$\\
  $^1$Dominion Astrophysical Observatory, Herzberg Institute of 
  Astrophysics, National Research Council of Canada, 5071 West Saanich Road, \\ 
  \hspace{5mm}Victoria, British Columbia, V9E 2E7, Canada. 
  \hspace{1mm} E-mail: sidney.vandenbergh@nrc-cnrc.gc.ca \\
  $^2$Institute of Astronomy, University of Cambridge, Madingley Road,
  Cambridge CB3 0HA, UK. \hspace{1mm} E-mail: dmackey@ast.cam.ac.uk}
\date{Accepted --. Received --}
\pagerange{000--000}
\pubyear{2004}

\def\LaTeX{L\kern-.36em\raise.3ex\hbox{a}\kern-.15em
    T\kern-.1667em\lower.7ex\hbox{E}\kern-.125emX}

\begin{document}

\label{firstpage}

\maketitle

\begin{abstract}
Globular clusters in the outer halo ($R_{\rm{gc}} > 15$ kpc) are found to be
systematically fainter than those at smaller Galactocentric distances.
Within the outer halo the compact clusters with half-light radii 
$R_h < 10$ pc are only found at $R_{\rm{gc}} < 40$ kpc, while on the other 
hand the larger clusters with $R_h > 10$ pc are encountered at all 
Galactocentric distances. Among the compact clusters with $R_h < 10$ pc 
that have $R_{\rm{gc}} > 15$ kpc, there are two objects with surprisingly 
high metallicities. One of these is Terzan 7, which is a companion of 
the Sagittarius dwarf. The other is Palomar 1. The data on these two 
objects suggests that they might have had similar evolutionary histories.
It is also noted that, with one exception, luminous globular clusters in 
the outer halo are all compact whereas faint ones may have any radius. 
This also holds for globular clusters in the LMC, SMC and Fornax dwarf. 
The lone exception is the large luminous globular NGC 2419. Possibly 
this object is not a normal globular cluster, but the stripped core of a 
former dwarf spheroidal. In this respect it may resemble $\omega$ Centauri.
\end{abstract}


\section{Introduction}
\label{s:intro}
Between 1962 and 1977 it was generally believed that the Galaxy had
formed by rapid dissipative collapse of a single massive proto-Galaxy.
Faith in this paradigm was severely shaken by two papers presented at
the 1977 Yale conference on {\it The Evolution of Galaxies and Stellar
Populations} \cite{tinsley:77}. In one of these Searle 
\shortcite{searle:77} showed that, contrary to expectations from the 
Eggen, Lynden-Bell \& Sandage \shortcite{eggen:62} (= ELS) model, 
globular clusters in the outer halo of the Galaxy did not exhibit a 
metallicity gradient (although we note that it is not necessarily expected 
that a radial abundance gradient be set up in the very outer part of the Galaxy 
if the initial phase of the ELS collapse is in free-fall \cite{sandage:90}).
Furthermore Toomre \shortcite{toomre:77} 
pointed out that {\it ``It seems almost inconceivable that there wasn't a great 
deal of merging of sizable bits and pieces (including quite a few lesser 
galaxies) early in the career of every major galaxy''.} 
These ideas were elaborated upon by Searle \& Zinn 
\shortcite{searle:78} (= SZ) who argued that the lack of an abundance gradient
in the outer Galactic halo, along with anomalies in the colour-magnitude diagrams of 
outer halo clusters suggested that the outer Galactic halo was assembled over an 
extended duration via the infall of transient proto-Galactic fragments.
Zinn \shortcite{zinn:80} describes this process of accretion of proto-Galactic 
gas clouds in more detail: {\it ``The clouds in the central zone [of the Galaxy] merged 
to form a large gas cloud that later evolved into the Galactic disk. The clouds in 
the second zone evolved as isolated systems for various lengths of time up to
$\mathit{\sim 5}$ Gyr, but eventually all of these clouds collided with the disk and were 
destroyed. The clouds of the outermost zone have evolved in relative isolation
until the present time and have become the Magellanic Clouds and the
dwarf spheroidal galaxies''.} For additional detailed information the reader is 
referred to van den Bergh \shortcite{vdb:95,vdb:04}.

Presently available data appear to favour the view that the main body of the Milky 
Way system, i.e. the region with $R_{\rm{gc}} < 10$ kpc, was formed more-or-less 
\'{a} la ELS (as modified by Sandage \shortcite{sandage:90}), whereas the region 
with $R_{\rm{gc}} > 15$ kpc was probably mainly assembled by infall and capture of 
lesser fragments as envisioned in the SZ model. Zinn \shortcite{zinn:93} first
pointed out that the observed data for the halo globular clusters was best explained
in terms of both ELS and SZ. He split the halo globular clusters into two groups
according to their horizontal branch morphologies. The properties of Zinn's ``old 
halo'' (blue horizontal branch) sub-system are consistent with the majority of its
members having been formed as part of an ELS-type collapse, while the properties of 
his ``young halo'' (red horizontal branch, or second parameter) sub-system are 
more in line with its members having been formed in ancestral fragments and accreted 
into the outer halo at later epochs. Additional support for this view is provided 
by the observation \cite{vdb:93} that a significant fraction of the globular 
clusters at $R_{\rm{gc}} > 15$ kpc are on plunging orbits, and that a number of 
outer halo objects with relatively well-determined orbits have retrograde motions
\cite{dinescu:99}. It is noted in passing \cite{vdb:95} that cluster metallicity 
appears to correlate more strongly with the peri-Galactic distances of globular 
clusters than it does with the present Galactic distance $R_{\rm{gc}}$.

\section{Globular clusters in the outer Galactic halo}
\label{s:outer}
In the present paper we use data in the recent compilation by
Harris \shortcite{harris:96} (2003 update) to study some of the 
properties of the globular clusters in the outer halo ($R_{\rm{gc}} > 15$ 
kpc) in an attempt to find out more about how the outer Galactic halo 
was assembled. Not all galaxies are formed in the same way. Whereas the 
main body of our own Milky Way system seems to have formed via the
early collapse of a single large protogalaxy the Andromeda galaxy (M31)
appears to have been assembled at a later date via the merger of at 
least two major ancestral fragments \cite{freeman:99,vdb:04,brown:04}.

\begin{table}
\begin{center}
\caption{Globular clusters with $R_{\rm{gc}} > 15$ kpc}
\begin{tabular}{@{}lccccc}
\hline \hline
Name & \hspace{10mm} & $R_{\rm{gc}}$ & $M_V$ & $[$Fe$/$H$]$ & \hspace{2mm} $R_h$ \hspace{2mm} \\
 & & (kpc) & & & (pc) \\
\hline
NGC 1261 &       &  $18.2$  & $-7.81$ &   $-1.35$ &  $3.6$     \\
Pal. 1   &       &  $17.0$  & $-2.47$ &   $-0.60$ &  $2.2$     \\
AM 1     &       &  $123.2$ & $-4.71$ &   $-1.80$ &  $17.7$    \\
Eridanus &       &  $95.2$  & $-5.14$ &   $-1.46$ &  $10.5$    \\
Pal. 2   &       &  $35.4^{*}$ & $-8.01$ &  $-1.30$ &  $5.4$   \\
NGC 1851 &       &  $16.7^{+}$  & $-8.33$ &   $-1.22$ &  $1.8$     \\
NGC 1904 &  M79  &  $18.8^{+}$  & $-7.86$ &   $-1.57$ &  $3.0$     \\
NGC 2298 &       &  $15.7^{+}$  & $-6.30$ &   $-1.85$ &  $2.4$     \\
NGC 2419 &       &  $91.5$  & $-9.58$ &   $-2.12$ &  $17.9$    \\
NGC 2808$^{1}$ &       &  $11.1^{+}$  & $-9.39$ &   $-1.15$ &  $2.1$     \\
Pyxis    &       &  $41.7$  & $-5.75$ &   $-1.30$ &  $15.6$    \\
Pal. 3   &       &  $95.9$  & $-5.70$ &   $-1.66$ &  $17.8$    \\
Pal. 4   &       &  $111.8$ & $-6.02$ &   $-1.48$ &  $17.2$    \\
NGC 4147 &       &  $21.3^{*}$ & $-6.16$ &   $-1.83$ &  $2.4$  \\
Rup. 106 &       &  $18.5$  & $-6.35$ &   $-1.67$ &  $6.8$     \\
NGC 5024 &  M53  &  $18.3$  & $-8.70$ &   $-1.99$ &  $5.7$     \\
NGC 5053 &       &  $16.9$  & $-6.72$ &   $-2.29$ &  $16.7$    \\
AM 4     &       &  $25.5$  & $-1.60$ &   $-2.00$ &  $3.7$     \\
NGC 5466 &       &  $16.2$  & $-6.96$ &   $-2.22$ &  $10.4$    \\
NGC 5634 &       &  $21.2$  & $-7.69$ &   $-1.88$ &  $4.0$     \\
NGC 5694 &       &  $29.1$  & $-7.81$ &   $-1.86$ &  $3.3$     \\
IC 4499  &       &  $15.7$  & $-7.33$ &   $-1.60$ &  $8.2$     \\
NGC 5824 &       &  $25.8$  & $-8.84$ &   $-1.85$ &  $3.4$     \\
Pal. 5   &       &  $18.6$  & $-5.17$ &   $-1.41$ &  $20.0$    \\
Pal. 14  &  AvdB &  $69.0$  & $-4.73$ &   $-1.52$ &  $24.7$    \\
NGC 6229 &       &  $29.7$  & $-8.05$ &   $-1.43$ &  $3.3$     \\
Pal. 15  &       &  $37.9$  & $-5.49$ &   $-1.90$ &  $15.7$    \\
IC 1257  &       &  $17.9$  & $-6.15$ &   $-1.70$ &  $...$     \\
NGC 6715 &  M54  &  $19.2^{*}$ & $-10.01$ &  $-1.58$ &  $3.8$  \\
Ter. 7   &       &  $16.0^{*}$ & $-5.05$ &   $-0.58$ &  $6.6$  \\
Arp 2    &       &  $21.4^{*}$ & $-5.29$ &   $-1.76$ &  $15.9$ \\
Ter. 8   &       &  $19.1^{*}$ & $-5.05$ &   $-2.00$ &  $7.6$  \\
NGC 7006 &       &  $38.8$  & $-7.68$ &   $-1.63$ &  $4.6$     \\
Pal. 12  &       &  $15.9^{*}$ & $-4.48$ &   $-0.94$ &  $7.1$  \\
Pal. 13  &       &  $26.7$  & $-3.74$ &   $-1.74$ &  $3.5$     \\
NGC 7492 &       &  $24.9$  & $-5.77$ &   $-1.51$ &  $9.2$     \\
\hline
\label{t:galactic}
\end{tabular}
\medskip
\begin{minipage}{80mm}
$^{*}$ Probably associated with the Sagittarius dwarf. \\
$^{+}$ Possibly associated with the disrupted Canis Major dwarf. \\
$^{1}$ NGC 2808 is included in this Table even though it has $R_{\rm{gc}} < 15$ kpc, 
because of its possible association with the Canis Major dwarf.
\end{minipage}
\end{center}
\end{table}

The Harris catalog contains $35$ Galactic clusters with $R_{\rm{gc}} > 15$ 
kpc. Data on these clusters are listed in Table \ref{t:galactic}. In this
Table we have also included data for NGC 2808, which has $R_{\rm{gc}} = 11.1$ kpc,
because of its possible association with the Canis Major dwarf (see below). 
Of the $35$ objects with $R_{\rm{gc}} > 15$ kpc, the following seven appear 
to be associated with the Sagittarius dwarf galaxy: M54, Terzan 7 and 8, and Arp 2 
\cite{ibata:94,dacosta:95}; Pal. 12 and NGC 4147 
\cite{dinescu:00,md:02,bellazzini:03a}; and Pal. 2 
\cite{majewski:04}. Also note the possible physical association of 
the globular clusters NGC 1851, NGC 1904, NGC 2298 and NGC 2808, plus a number 
of old open clusters, with the recently discovered Canis Major dwarf 
\cite{martin:04,bellazzini:03b,frinchaboy:04}.
In particular, Bellazzini et al \shortcite{bellazzini:03b} find strong
evidence that the clusters AM 2 and Tombaugh 2 [which are not cataloged as 
globulars by Harris] are associated with the Canis Major system. In fact, 
these authors suggest that Tombaugh 2 may actually represent an over-density 
in the CMa field itself, similar to those observed in the Sagittarius and Ursa 
Minor dwarf galaxies. Finally, Carraro et al. \shortcite{carraro:04} have shown that
the cluster Berkeley 29 is associated with the Monoceros stream, which is
thought \cite{martin:04} to be part of the disrupted CMa dwarf. 
However, this cluster has an age of $\sim 5$ Gyr, which makes it somewhat too 
young to be of interest for the present study of globular clusters.

Information on the globular clusters in the LMC, the SMC, and the
Fornax dwarf spheroidal is collected in Table \ref{t:external} using the
data from Mackey \& Gilmore \shortcite{mackey:03a,mackey:03b,mackey:03c}. 
The total luminosities and half-light radii in this Table have been newly 
derived for the present work. The total luminosities ($M_V$) were obtained 
by integrating these authors' radial brightness profiles to appropriate 
limiting radii ($\sim 50$ pc) using Eq. $12$ of Mackey \& Gilmore 
\shortcite{mackey:03a}. Rearranging this equation then allows the subsequent 
determination of the half-light radii. 
Distance moduli of $18.50$, $18.90$, and $20.68$ have been adopted for the 
LMC, SMC, and Fornax systems, respectively.

The LMC sample of Mackey \& Gilmore \shortcite{mackey:03a} omits four 
known globulars (NGC 1928, 1939, Reticulum, and ESO121-SC03). A recent 
{\em Hubble Space Telescope} program has obtained images of these four 
objects using the Advanced Camera for Surveys (e.g., Mackey \& Gilmore 
2004). Preliminary radial luminosity profiles (for which details will follow
in a future work -- Mackey \& Gilmore, in prep.) have allowed integrated 
magnitudes and half light radii to be estimated for these four clusters; 
however those for NGC 1928 and 1939 are very uncertain due to severe 
crowding in the cluster images. Half light radii for these two objects 
should be considered upper limits only.
   
The LMC cluster sample is therefore complete, consisting of all $16$
known globular cluster-type objects. The Fornax cluster sample is also
complete ($5$ clusters). For the SMC, we have only listed NGC 121. 
Although there are other reasonably old clusters in this galaxy (e.g., 
Kron 3, Lindsay 1), it is not clear that they are directly comparable to
the globular clusters in the Galactic halo.

It is the aim of the present paper to see if such data on the globular
clusters in nearby dwarf galaxies can provide us with hints about the
evolutionary history of the outer regions of our own Milky Way system.

\begin{table}
\begin{center}
\caption{Information on the globular clusters associated with
the Large Magellanic Cloud, the Small Magellanic Cloud, and the
Fornax dwarf.}
\begin{tabular}{@{}lccccc}
\hline \hline
Name & Galaxy & $M_V$ & $[$Fe$/$H$]$ & $R_h$ & $R_{\rm{LMC}}$ \\
 & & & & (pc) & ($\degr$) \\
\hline
NGC 1466    &  LMC   &    $-7.26$  &   $-2.17$  &   $4.8$    &    $8.4$    \\
NGC 1754    &  LMC   &    $-7.09$  &   $-1.54$  &   $3.2$    &    $2.6$    \\
NGC 1786    &  LMC   &    $-7.70$  &   $-1.87$  &   $3.3$    &    $2.5$    \\
NGC 1835    &  LMC   &    $-8.30$  &   $-1.79$  &   $2.4$    &    $1.4$    \\
NGC 1841    &  LMC   &    $-6.82$  &   $-2.11$  &   $10.8$   &    $14.9$    \\
NGC 1898    &  LMC   &    $-7.49$  &   $-1.37$  &   $8.4$    &    $0.6$    \\
NGC 1916    &  LMC   &    $-8.24$  &   $-2.08$  &   $2.2$    &    $0.2$    \\
NGC 1928    &  LMC   &    $-6.06${\bf :} &   $-1.27$  &   $5.6${\bf :}   &    $0.2$    \\
NGC 1939    &  LMC   &    $-6.85${\bf :} &   $-2.10$  &   $7.6${\bf :}   &    $0.7$    \\
NGC 2005    &  LMC   &    $-7.40$  &   $-1.92$  &   $2.7$    &    $0.9$    \\
NGC 2019    &  LMC   &    $-7.75$  &   $-1.81$  &   $2.9$    &    $1.3$    \\
NGC 2210    &  LMC    &   $-7.51$  &   $-1.97$  &   $3.5$    &    $4.4$    \\
NGC 2257    &  LMC   &    $-7.25$  &   $-1.63$  &   $10.5$   &    $8.4$    \\
Hodge 11    &  LMC   &    $-7.45$  &   $-2.06$  &   $8.6$    &    $4.7$    \\
Reticulum   &  LMC   &    $-5.22$  &   $-1.66$  &   $19.3$   &    $11.4$    \\
ESO121-SC03 &  LMC   &    $-4.37$  &   $-0.93$  &   $10.0$   &    $9.7$    \\
\hline
NGC 121     &  SMC   &    $-7.89$  &   $-1.71$  &   $5.4$     &   $...$    \\
\hline
Fornax 1    &  Fnx   &    $-5.32$  &   $-2.05$  &   $11.8$   &    $...$    \\
Fornax 2    &  Fnx   &    $-7.03$  &   $-1.83$  &   $8.2$    &    $...$    \\
Fornax 3    &  Fnx   &    $-7.66$  &   $-2.04$  &   $4.4$    &    $...$    \\
Fornax 4    &  Fnx   &    $-6.83$  &   $-1.95$  &   $3.5$    &    $...$    \\
Fornax 5    &  Fnx   &    $-6.82$  &   $-1.90$  &   $4.4$    &    $...$    \\
\hline
\label{t:external}
\end{tabular}
\end{center}
\end{table}

\subsection{The luminosities of outer halo globular clusters}
\label{ss:lumin}
\begin{figure}
\includegraphics[width=0.5\textwidth]{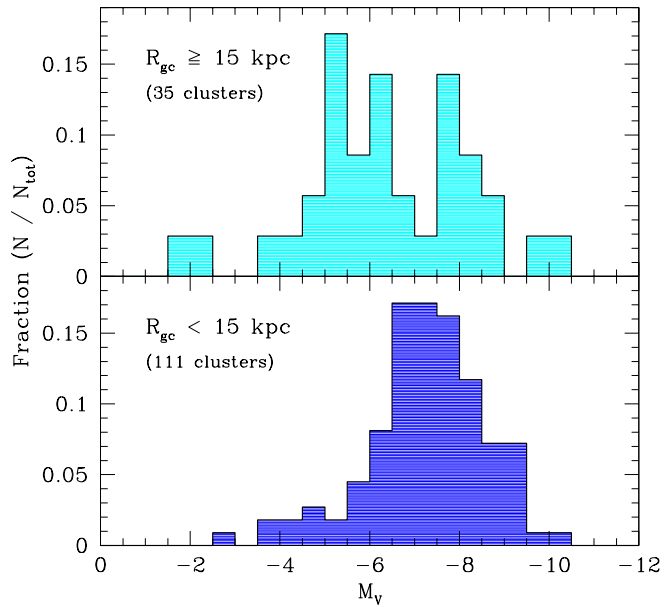}
\caption{Luminosity distribution of Galactic globular clusters
with $R_{\rm{gc}} < 15$ kpc (lower panel) and with $R_{\rm{gc}} \ge 15$ kpc (upper panel).
The observed systematic difference might, at least in part,
be due to the fact that bulge and disk shocks will preferentially
destroy low-mass clusters.}
\label{f:luminhist}
\end{figure}

It has been known for many years (e.g. van den Bergh 2000) that
globular clusters in the outer halo of the Galaxy are, on average,
fainter than those in the inner halo. This effect is shown in Figure 
\ref{f:luminhist}. A Kolmogorov-Smirnov test shows that there is only a 
$0.4$ per cent probability that the $M_V$ values of the $35$ globulars 
with $R_{\rm{gc}} \ge 15$ kpc and the $111$ globulars having $R_{\rm{gc}} < 15$ 
kpc listed in the updated version of Harris \shortcite{harris:96} were 
drawn from the same parent luminosity distribution.

This difference in luminosity between clusters at large and small
values of $R_{\rm{gc}}$ is, at least in part, due to the fact that globulars
at large $R_{\rm{gc}}$ values are less likely to have been destroyed by bulge
shocks \cite{aguilar:88} and tidal stresses \cite{surdin:94}
than clusters at smaller Galactocentric distances. It is noted in passing
that a Kolmogorov-Smirnov test shows no statistically significant
difference between the luminosity distribution of the seven clusters
likely to be associated with the Sagittarius dwarf and that of
the other $28$ clusters at $R_{\rm{gc}} \ge 15$ kpc. This is exactly what
one would expect if the Galactic outer halo globular clusters had
originally all been associated with dwarf spheroidal galaxies -- most of
which have since been destroyed. We point out, however, that the number
of clusters involved is too small to attach great significance to
this result.

Similarly, it is interesting to note that the four clusters which have been
suggested as members of the disrupted Canis Major dwarf are, on average, more 
luminous than the seven Sagittarius clusters and five Fornax clusters listed
in Tables \ref{t:galactic} and \ref{t:external}. However, both the difference 
and the number of clusters concerned are too small to constitute a convincing 
argument against the notion that these four clusters were originally formed
in a dSph-type galaxy.

It is also of interest to note that there is, perhaps, a hint of a 
relationship between luminosity and galactocentric radius among the
globular clusters in the Large Magellanic Cloud. All $13$ LMC clusters
with $R_{\rm{LMC}} < 10\degr$ are brighter than $M_V = -6.0$. On the other 
hand two of the three Large Cloud clusters with $R_{\rm{LMC}} \ge 10\degr$ 
are fainter than $M_V = -6.0$.

The existence of a metallicity gradient for globulars with 
$4 < R_{\rm{gc}} < 10$ kpc, and the absence of such a gradient for globular 
clusters with $R_{\rm{gc}} > 10$ kpc are consistent with the view 
(e.g. van den Bergh 2004) that the inner region of the Milky Way system 
formed by a collapse of the type envisioned by Eggen, Lynden-Bell \& 
Sandage \shortcite{eggen:62}, whereas the outer part of our Galaxy was 
mainly assembled by capture of bits and pieces in the manner suggested 
by Searle \& Zinn \shortcite{searle:78}.

\subsection{Radial luminosity dependence of globular clusters in the 
Galactic halo}
\label{ss:radlumin}
\begin{figure}
\includegraphics[width=0.5\textwidth]{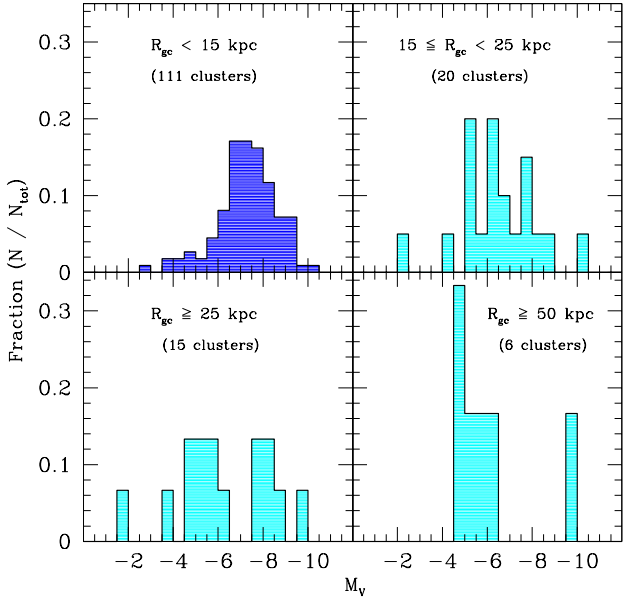}
\caption{Luminosity distributions of Galactic globular clusters
with $R_{\rm{gc}} < 15$ kpc (upper left) and with $15 \le R_{\rm{gc}} < 25$ kpc
(upper right); $R_{\rm{gc}} \ge 25$ kpc (lower left); and $R_{\rm{gc}} \ge 50$ kpc
(lower right). The samples appear systematically less luminous at
greater Galactocentric radii.}
\label{f:radluminhist}
\end{figure}

A comparison between the luminosity distributions of globular clusters
with $R_{\rm{gc}} < 15$ kpc and those with $15 \le R_{\rm{gc}} < 25$ kpc, as 
depicted in Figure \ref{f:radluminhist}, shows a tendency for the more 
distant cluster sample to be less luminous than the the inner sample. 
However, a Kolmogorov-Smirnov test shows that this difference is only 
significant at the $90$ per cent level. The difference between the 
luminosity distributions of the globular clusters with $R_{\rm{gc}} < 15$ 
kpc and those with $R_{\rm{gc}} \ge 25$ kpc is, however, much more striking, 
with faint clusters being deficient in the sample with small $R_{\rm{gc}}$. 
A Kolmogorov-Smirnov test shows that there is only a $1.5$ per cent 
probability that the samples with $R_{\rm{gc}} < 15$ kpc and with 
$R_{\rm{gc}} \ge 25$ kpc were drawn from the same parent population of cluster 
luminosities.

The most striking difference is seen between the clusters at 
$R_{\rm{gc}} \ge 50$ kpc and those at smaller Galactocentric radii. In this 
most distant sample only one cluster (NGC 2419) is brighter than 
$M_V = -6.1$. Perhaps the latter object is, like NGC 6715 = M54 
(e.g., Layden \& Sarajedini 2000) and $\omega$ Centauri (see e.g., 
Bekki \& Freeman 2003; Tsuchiya et al. 2003, and  references therein), 
the stripped remnant of the core of a former dwarf spheroidal companion 
to the Galaxy. If this hypothesis is correct one might expect to observe 
a significant scatter in the metallicities of stars on the giant branch 
of NGC 2419. However, the rather thinly populated color-magnitude 
diagram of NGC 2419 published by Harris et al. \shortcite{harris:97}
appears to show no evidence for such a scatter in metallicity. The 
apparent absence of such scatter in the color-magnitude diagram of 
NGC 2419 might be due to the fact that both the core and the spheroidal 
envelope of this object consist of very metal-poor stars. In other 
words the parent galaxy of NGC 2419 might, like the Ursa Minor 
dwarf, have produced only a single generation of quite metal-poor stars.

\subsection{Half-light radii of outer halo globular clusters}
\label{ss:rhouter}
\begin{figure}
\includegraphics[width=0.5\textwidth]{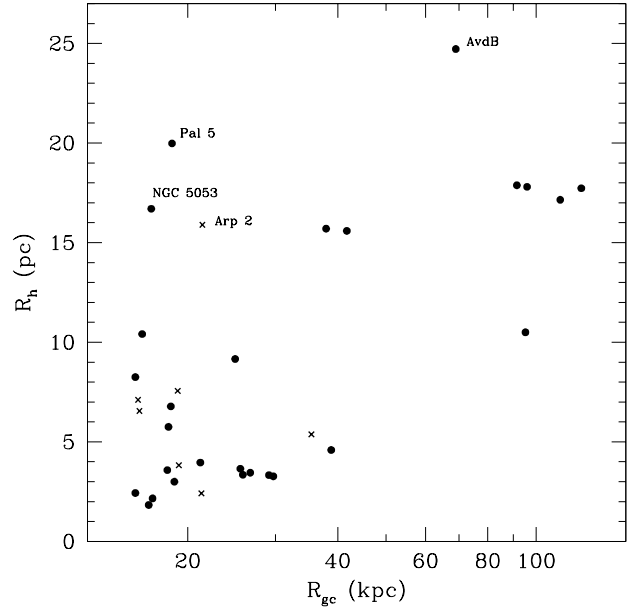}
\caption{Half-light radius $R_h$ versus Galactocentric distance
$R_{\rm{gc}}$ for Galactic globular clusters with $R_{\rm{gc}} > 15$ kpc (dots).
Also plotted are the globular clusters associated with the Sagittarius
dwarf galaxy (x marks). The figure shows that compact
clusters do not occur at $R_{\rm{gc}} > 40$ kpc.}
\label{f:rhvsrg}
\end{figure}

Figure \ref{f:rhvsrg} shows a plot of the dependence of the cluster 
half-light radius $R_h$, which does not change much due to dynamical 
evolution \cite{spitzer:72,lightman:78,murphy:90}, on Galactocentric 
distance $R_{\rm{gc}}$. The figure shows that compact clusters (defined as 
objects having $R_h < 10$ pc) only occur at $R_{\rm{gc}} < 40$ kpc, whereas 
more extended clusters with $R_h > 10$ pc are found to be located at 
all radii. However, it seems likely that a few very extended
clusters at relatively small Galactocentric distances, such as NGC 5053,
might actually be outer halo clusters that are presently orbiting closer
to the Galactic center. Odenkirchen et al. \shortcite{odenkirchen:03} 
have shown that the outer halo cluster Pal. 5 is very close to complete 
tidal disruption, a fact which might help explain its large half-light radius
We note that it is not clear whether Pal. 5 has a large radius entirely because 
of tidal effects, or was it actually a large initial radius that allowed tidal 
effects to become important. The most extended globular presently known is 
Pal. 14 = AvdB, which has a half-light radius of $24.7$ pc. It is of interest to 
note that the cluster Arp 2, which appears to be associated with the Sagittarius 
system, has a large half-light radius $R_h = 15.9$ kpc, but that the 
other globulars associated with the Sagittarius system are all compact.
   
It is of interest to point out that Forbes, Strader \& Brodie \shortcite{forbes:04},
in their study of the prospective cluster systems of the Canis Major and Sagittarius
dwarf galaxies, found that the four CMa clusters (NGC 1851, 1904, 2298, and 2808)
have unusually small radii for their Galactocentric distance.

\begin{figure}
\includegraphics[width=0.5\textwidth]{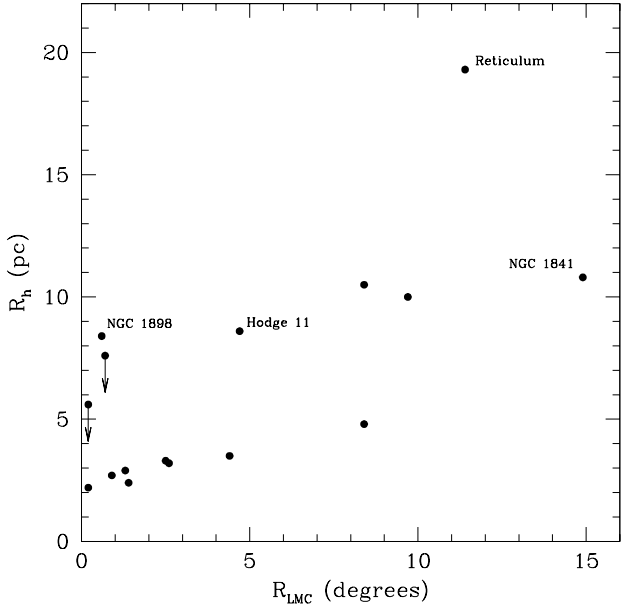}
\caption{Half-light radius $R_h$ versus $R_{\rm{LMC}}$, the angular distance
from the rotation centre of the LMC, for $16$ globular clusters in the
Large Cloud. We calculate $R_{\rm{LMC}}$ relative to the LMC rotation centre
at $\alpha = 05^{\rm{h}} 20^{\rm{m}} 40^{\rm{s}}$, $\delta = -69\degr 14\arcmin 12\arcsec$
(J2000.0).}
\label{f:rlmc}
\end{figure}

It has been known for many years (e.g. van den Bergh \& Morbey 1984)
that the half-light radii of Galactic globular clusters grow with
increasing Galactocentric distance. By the same token it has long
been known \cite{hodge:62} that old clusters near the core of the LMC
are generally more compact than are clusters at larger distances from
the center of the Large Cloud. Using our new data on cluster half-light
radii this effect is clearly shown in Figure \ref{f:rlmc}. This Figure
shows that the majority of those clusters close to the rotation centre
of the LMC are compact, while those at large distances are much more 
extended. The largest LMC cluster, Reticulum, is one of the most distant;
while the most distant cluster, NGC 1841, is one of the most extended.
The clusters NGC 1898 and Hodge 11 are larger than might be expected
given their comparatively small distances from the LMC centre.
These objects are possibly LMC halo clusters projected
on the inner parts of this galaxy.

Finally, we note that extended globular clusters have a luminosity function which
differs strongly from that of more compact globular clusters (e.g., van den Bergh
\shortcite{vdb:82}). It seems likely that this is related to the observed correlations
between cluster luminosity and Galactocentric radius (see the previous Section),
and cluster size and and Galactocentric radius (this Section).

\subsection{Metallicity and cluster size}
\label{ss:metsize}
\begin{figure}
\includegraphics[width=0.5\textwidth]{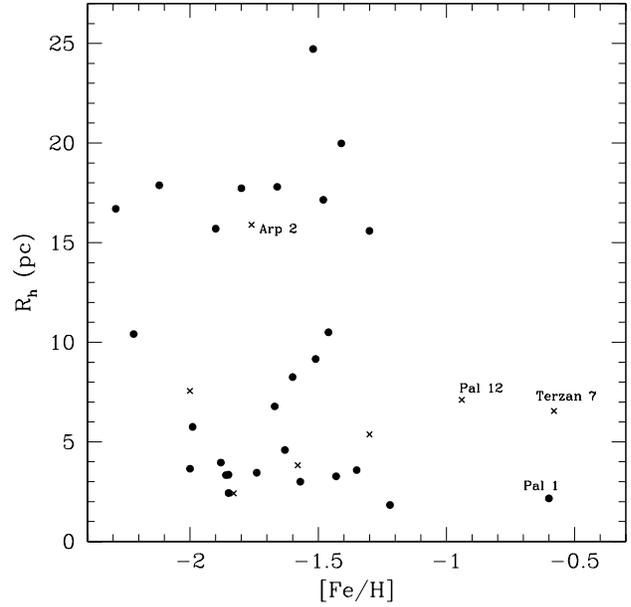}
\caption{Half-light radius $R_h$ versus metallicity $[$Fe$/$H$]$. The
proximity of Ter. 7 and Pal. 1 suggests that these objects
might have had similar evolutionary histories. Pal. 12 also
falls close to these two objects. Galactic halo clusters are
shown as dots and Sagittarius globulars as crosses.}
\label{f:rhvsmet}
\end{figure}

A plot of half-light radius $R_h$ versus metallicity $[$Fe$/$H$]$ is 
shown in Figure \ref{f:rhvsmet}. Most of the globular clusters associated
with the Sagittarius dwarf \cite{ibata:94,forbes:04}, which are plotted 
as crosses in Fig. \ref{f:rhvsmet}, have small $R_h$ values.
However the large cluster Arp 2, with $R_h = 15.9$ pc, is an
exception to this rule. Among the clusters with $R_{\rm{gc}} > 15$ kpc there
are two that are surprisingly metal rich. One of these is Terzan 7,
which is believed to be associated with the Sagittarius dwarf, whereas
the other (Palomar 1) appears to be an isolated object. Nevertheless
the unusually high metallicities of these two objects suggest that they
could have had similar evolutionary histories, i.e. Pal. 1 might once
have been associated with a dwarf spheroidal galaxy that was subsequently
destroyed by tidal forces. It is also noted that the cluster Palomar 12,
which is associated with the Sagittarius dwarf, occurs close to the
region of the $R_h$ vs. $[$Fe$/$H$]$ diagram in which Ter. 7 and Pal. 1 
are located.
   
Finally (and not unexpectedly) the most distended outer halo clusters
generally have lower metallicities than the more compact objects that
typically inhabit the zone with small $R_{\rm{gc}}$ values.

\subsection{Cluster radius and cluster luminosity}
\label{ss:luminrad}

\begin{figure}
\includegraphics[width=0.5\textwidth]{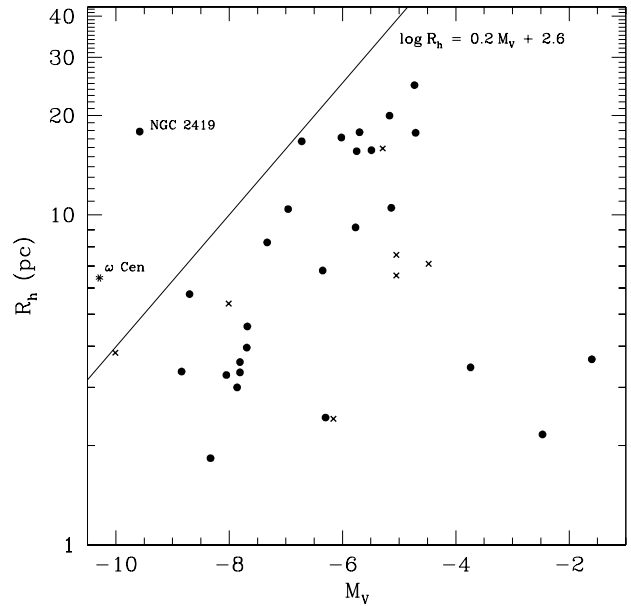}
\caption{Half-light radius $R_h$ versus luminosity $M_V$ of globular
clusters with $R_{\rm{gc}} > 15$ kpc. The figure shows that all outer
globulars, except NGC 2419, lie below the relation given
by Eq. \ref{e:cutoff} which is shown as a slanting line in the Figure.
$\omega$ Centauri (plotted as a star) also lies significantly above
this relation. Clusters associated with the Sagittarius dwarf are
shown as crosses.}
\label{f:rhvsmv}
\end{figure}

\begin{figure}
\includegraphics[width=0.5\textwidth]{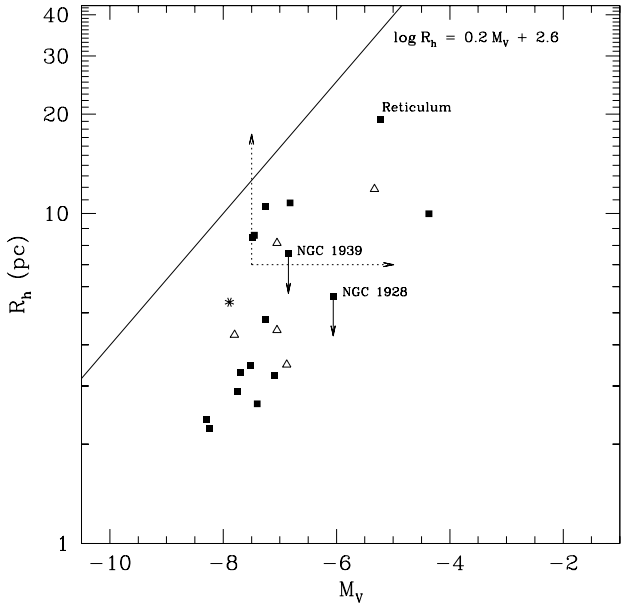}
\caption{Half-light radius $R_h$ versus luminosity $M_V$ for globular
clusters associated with the LMC (filled squares), the SMC (star)
and Fornax dwarf (open triangles). Together with Figure \ref{f:rhvsmv}, this plot shows
that the globular clusters in the outer Galactic halo having $R_{\rm{gc}} > 15$
kpc occupy the same region in the diagram as do the globular clusters
in the LMC, the SMC, and the Fornax and Sagittarius dwarfs. The ``faint fuzzies'' 
of Larsen \& Brodie (2000) lie above and to the right of the
two dotted lines (Brodie \& Larsen 2003). This shows that, in purely structural terms, these
objects are not dissimilar to the most extended clusters in the outer
Galactic halo and nearby dwarf galaxies.}
\label{f:rhvsmvexternal}
\end{figure}

Figure \ref{f:rhvsmv} shows that the halo globular clusters with 
$R_h > 15$ kpc do not fill the $R_h$ vs $M_V$ plane uniformly. No 
clusters, except NGC 2419, lie above the relation
\begin{equation}
\log R_h = 0.2 M_V + 2.6 .
\label{e:cutoff}
\end{equation}
The figure shows that a compact halo globular cluster can have any
luminosity, while large globular clusters in the outer halo can only be
intrinsically faint. Furthermore, the most luminous globular clusters tend
to be among the most compact. The only exception to this rule is NGC 2419. As
noted in Section \ref{ss:radlumin}, it is possible that this large 
luminous object is actually the surviving core of a dwarf spheroidal 
galaxy, rather than a normal globular cluster. The atypical inner halo 
cluster $\omega$ Centauri is also marked on Figure \ref{f:rhvsmv}. Like 
NGC 2419, it lies significantly above the line defined by Eq. 
\ref{e:cutoff}. There are many suggestions in the literature (see e.g., 
Bekki \& Freeman 2003, and the references therein) that $\omega$ Centauri is 
the remaining core of a now disrupted dwarf galaxy. The position of this 
cluster in Figure \ref{f:rhvsmv} adds plausibility to our suggestion 
that NGC 2419 is of a similar nature.
   
It is noted in passing that the globular clusters associated with the
Sagittarius dwarf, which are plotted as crosses in Figure \ref{f:rhvsmv},
appear to be distributed similarly to other globular clusters at 
$R_{\rm{gc}} > 15$ kpc.

Figure \ref{f:rhvsmvexternal} shows a plot of the distribution of the 
globular clusters in the LMC, the SMC and the Fornax dwarf in the 
$R_h$ vs. $M_V$ plane. This figure shows that these external globular 
clusters also fall below the line defined by Eq. \ref{e:cutoff}, 
inhabiting a comparable region of the plane to that occupied by the 
globular clusters in the outer Galactic halo. The LMC Reticulum cluster 
is the most extended object in the external sample, with $R_h = 19.3$ pc.
The resemblance between the external globular clusters and those in the 
outer Galactic halo may suggest a similar origin -- i.e., in dwarf 
spheroidal-like galaxies.
   
It is presently not clear why the most luminous, and hence presumably
most massive, globular clusters have the smallest radii. This conclusion
appears to hold true both for Galactic globular clusters and for those
associated with the Magellanic Clouds and the dwarf spheroidal
companions of the Milky Way. It is noted that in the case of a constant
cluster mass-to-light ratio, Eq. \ref{e:cutoff} implies that the upper
limit to globular cluster sizes in the outer halo (excluding NGC 2419) is
defined by
\begin{equation}
M R_h^2 = {\rm const} ,
\label{e:mrh}
\end{equation}
where $M$ is cluster mass. In form, this is perhaps reminiscent to the 
correlation $R_h \propto M^{-0.63}$ noted by Ostriker \& Gnedin 
\shortcite{ostriker:97} for clusters with $5 < R_{\rm{gc}} < 60$ kpc.
The majority of clusters in Figures \ref{f:rhvsmv} and \ref{f:rhvsmvexternal}
appear to follow a similar correlation, running approximately parallel to 
the line defined by Eq. \ref{e:cutoff}.

McLaughlin \shortcite{mclaughlin:00} defines a fundamental plane for 
globular clusters using observations of clusters at all Galactocentric
radii. Although he only finds a very weak correlation between luminosity
(mass) and half-light radius, it seems plausible that the cut-off we
observe in the present work is related to the presence of a fundamental
plane for globular clusters. McLaughlin argues that the characteristics
of the fundamental plane were set by the cluster formation process.
The fact that NGC 2419 and $\omega$ Centauri fall above our cut-off and away
from all other globular clusters seems likely to place them away
from the fundamental plane. If this hypothesis is correct then it would
imply a formation scenario for NGC 2419 and $\omega$ Centauri different from
that for the rest of the Galactic globular clusters. One such scenario is
that these two objects are not true globular clusters, but rather the
remaining cores of now defunct dwarf galaxies.
   
Finally, it is informative to place the ``faint fuzzy'' clusters
discovered in the lenticular galaxies NGC 1023 and 3384 by Larsen \& Brodie 
\shortcite{larsen:00} and Larsen et al. \shortcite{larsen:01} on our 
$\log R_h$ vs. $M_V$ plot. In their Figures 1 and 2, Brodie \& Larsen 
\shortcite{brodie:03} show that the faint fuzzies have $R_h \ge 7$ pc 
and integrated luminosities fainter than $M_V = -7.5$.
These define a region on our Figure \ref{f:rhvsmvexternal}, marked by 
two dotted lines. The position of this region shows that while the faint 
fuzzies are clearly very distinct from any globular clusters in the local
neighbourhood in terms of their composition and spatial and kinematic properties 
\cite{brodie:03}, they are not so dissimilar purely in terms of 
luminosity and structure to the most extended clusters in the outer 
Galactic halo and nearby dwarf galaxies. It would be important to know 
if any of the largest faint fuzzies lie above our Eq. \ref{e:cutoff}, as 
appears possible from Figure \ref{f:rhvsmvexternal}.

\section{Conclusions}
\label{s:conclusions}
Presently available observations appear to be consistent with the
hypothesis that the globular clusters in the outer ($R_{\rm{gc}} > 15$ kpc) 
Galactic halo were once all associated with dwarf spheroidal-like
fragments that have since disintegrated. On the other hand the majority
of inner halo globular clusters with $R_{\rm{gc}} < 10$ kpc were probably 
formed in association with the main body of the Milky Way system. The 
fact that the sample of outer halo globulars contains more faint clusters
than the inner halo is likely due to the destruction of low-mass
inner clusters by disk shocks and tidal stripping. By the same token the
scarcity of large globulars at small $R_{\rm{gc}}$ values is likely also due to
such destructive forces. The presence of a few quite metal-rich clusters,
such as Terzan 7 ($[$Fe$/$H$] = -0.58$) and Palomar 1 
($[$Fe$/$H$] = -0.60$), and perhaps Palomar 12 ($[$Fe$/$H$] = -0.98$) at 
quite large Galactocentric radii appears anomalous. The 
existence of such metal-rich objects in the outer Galactic halo can 
be explained if they formed in, or in association with, dwarf spheroidal 
galaxies (as appears likely for Ter. 7 and Pal. 12). With one exception, 
luminous globular clusters in the outer 
halo are all compact whereas faint ones may have any radius, a result 
which also holds for globular clusters in the LMC, SMC and Fornax dwarf 
spheroidal. We speculate that the luminous ($M_V = -9.58$) and very 
large ($R_h = 17.9$ pc) cluster NGC 2419, which is located at a 
Galactocentric distance of $91.5$ kpc, might be the remnant core of a 
now dispersed dwarf spheroidal galaxy. Apart from apparently not 
possessing an internal metallicity spread, its properties are similar 
to the very large and luminous globular cluster $\omega$ Centauri, which 
might also be such a stripped core of a former dwarf spheroidal galaxy.

\section*{Acknowledgements}
We thank Paul Hodge for reminding us of his prescient 1962 paper on
the diameter and structure of clusters in the LMC. ADM recognises
financial support from PPARC in the form of a Postdoctoral Fellowship,
and is grateful to Mark Wilkinson for a useful discussion.



\bsp 

\label{lastpage}

\end{document}